# FALSE DISCOVERY RATE ANALYSIS OF BRAIN DIFFUSION DIRECTION MAPS

By Armin Schwartzman,[1] Robert F. Dougherty[2]
and Jonathan E. Taylor[3]

*Harvard School of Public Health, Stanford University
and Stanford University*

Diffusion tensor imaging (DTI) is a novel modality of magnetic resonance imaging that allows noninvasive mapping of the brain's white matter. A particular map derived from DTI measurements is a map of water principal diffusion directions, which are proxies for neural fiber directions. We consider a study in which diffusion direction maps were acquired for two groups of subjects. The objective of the analysis is to find regions of the brain in which the corresponding diffusion directions differ between the groups. This is attained by first computing a test statistic for the difference in direction at every brain location using a Watson model for directional data. Interesting locations are subsequently selected with control of the false discovery rate. More accurate modeling of the null distribution is obtained using an empirical null density based on the empirical distribution of the test statistics across the brain. Further, substantial improvements in power are achieved by local spatial averaging of the test statistic map. Although the focus is on one particular study and imaging technology, the proposed inference methods can be applied to other large scale simultaneous hypothesis testing problems with a continuous underlying spatial structure.

**1. Introduction.** A central statistical problem in brain imaging studies is to find areas of the brain that differ between two groups of subjects, namely, a control group and another group with a special characteristic of interest. Diffusion tensor imaging (DTI) is a modality of MRI that allows insight into

Received April 2007; revised April 2007.
[1]Supported in part by a William R. and Sara Hart Kimball Stanford Graduate Fellowship.
[2]Supported by the Schwab Foundation for Learning and NIH Grant EY-01-5000.
[3]Supported by NSF Grant DMS-04-05970.
*Key words and phrases.* Diffusion tensor imaging, directional statistics, multiple testing, empirical null, spatial smoothing.

the brain's white matter. As opposed to functional MRI, which shows brain activity, DTI reveals anatomical structure. DTI measures the diffusion of water molecules in tissue [Bammer et al. (2002), Basser and Pierpaoli (1996) and Le Bihan et al. (2001)]. Since the movement of water is affected by the cell structure, the pattern of diffusion is an indicator of the microscopic properties of the tissue. DTI data differ fundamentally from conventional imaging data in that values at each spatial location are not scalars but $3 \times 3$ positive definite matrices, also called diffusion tensors (DTs). The DT at a location in space can be thought of as the covariance matrix of a 3D Gaussian distribution that models the local Brownian motion of the water molecules in that location. The DTs are measured at discrete volume elements called voxels arranged in a regular spatial grid. Voxels are typically about 2 mm in size. A typical DT image of the entire brain may contain a few hundred thousand voxels.

Because of familiarity with scalar statistics, investigators frequently restrict their analysis of DTI data to scalar quantities derived from the DT [e.g., Bammer et al. (2000) and Deutsch et al. (2005)]. The two most important such quantites are trace and fractional anisotropy (FA), both functions of the DT's eigenvalues related respectively to the total amount of diffusion and the degree of anisotropy within a voxel. The most important nonscalar quantity derived from the DT is the principal diffusion direction, defined as the eigenvector corresponding to the largest eigenvalue of the DT. It is generally assumed that diffusion is restricted in the direction perpendicular to the nerve fibers, so the principal diffusion direction provides a proxy for the fiber direction within the voxel [Le Bihan et al. (2001)]. Thus, inference about the principal diffusion direction is valuable for understanding where in the brain different neural fibers may be directed to. The other two eigenvectors are not as easily interpretable. Fortunately, there is a rich literature in directional statistics that can help with this problem. To the best of our knowledge, the first attempt to formally analyze principal diffusion direction maps in a multi-subject study has been reported by the authors [Schwartzman et al. (2005)]. The present article is an extension of the analysis reported there.

Given two sets of principal diffusion direction maps, the task at hand is to find regions that differ between the two groups. After appropriate spatial registration so that all images are aligned in the same coordinate system, the analysis can be broken down into two main steps: (1) Computation of a statistic that tests the difference between group mean directions at every voxel; (2) Inference on the overall test statistic map.

For step one, we use a probability model for the principal diffusion direction given by the bipolar Watson distribution on the unit sphere [Best and Fisher (1986), Mardia and Jupp (2000) and Watson (1965)]. We chose this distribution because it is one of the simplest that possesses the property of being antipodally symmetric, giving to each direction and its negative the same probability. This is crucial because the DT is invariant under sign changes of the principal eigenvector. This particular model leads to appropriate definitions of mean direction and dispersion for a random sample of directions, as well as a test statistic for testing whether two samples of directions, that is, same voxel across two groups of subjects, have the same mean. According to this model, the test statistic under the null hypothesis has an approximate $F$ distribution for fixed sample size, asymptotically as the samples become more concentrated around their mean. This is useful when the number of subjects is small, as it is generally the case in imaging studies.

The second step in the analysis corresponds to solving a multiple testing problem among a large number of voxels. For this we use the procedure by Storey, Taylor and Siegmund (2004) that controls the false discovery rate (FDR). FDR inference has been applied successfully to microarray analysis [e.g., Efron et al. (2001) and Efron (2004)], but is still a relatively new technique in brain imaging [Genovese, Lazar and Nichols (2002) and Pacifico et al. (2004)], where, in contrast, the usual approach has been control of the family-wise error rate (FWER) using Gaussian random field theory [Worsley et al. (1996)]. A practical reason for our choosing FDR over FWER is that the Storey, Taylor and Siegmund (2004) procedure depends only on the marginal distribution of the test statistics, while FWER control would require knowledge of the covariance properties of the random field defined by the test statistics. This is relevant because the marginal distribution of the Watson statistic is easy to estimate, while the field properties are not. A methodological reason for using FDR is that sometimes researchers are not so interested in controlling the error over the entire search region but rather finding interesting regions that could be further investigated. For this reason it is appropriate to use the convention introduced by Efron (2004) of calling the selected voxels "interesting."

The innovative aspects of the statistical analysis are twofold. The first is a new empirical null for global modeling of the test statistic. Since the number of subjects is small, a parsimonious model (such as the Watson) is needed at each voxel. On the other hand, the number of voxels is large, so more accurate modeling of the null distribution of the test statistic can be obtained by considering global parameters common to all voxels. These global parameters are fit based on the empirical distribution of the test statistic among voxels that may be considered to belong to the null class. This empirical null concept was originally suggested for $z$-scores in a microarray experiment setting [Efron (2004)]. We introduce here a new version of the empirical null adapted to the $F$ nature of the Watson test statistic.

The second innovative aspect is the increase of power in FDR inference through spatial smoothing. An important distinction between the multiple testing problems in brain imaging and microarray analysis is that the former are accompanied by a spatial structure. We take advantage of this property to increase statistical power by applying local spatial averaging, which reduces the noise variance. While routinely used in image processing, the effect of spatial smoothing on FDR inference is only starting to be studied [Pacifico et al. (2004)]. Here the empirical null helps again assess the marginal null distribution of the test statistic after smoothing.

The article is organized as follows. Section 2 describes the data. Section 3, together with the Appendix, summarize the relevant features of the Watson distribution. Section 4 describes the FDR inference, including the empirical null and the local averaging. The data analysis is integrated into Sections 3 and 4 and follows along with the theory. Further evaluation and criticism of the analysis results are offered in Section 5.

**2. The data.** Our particular dataset concerns an observational study of anatomical differences related to reading ability in children, conducted by a research team at Stanford University that included one of the authors (RFD). The study was motivated by a previous report of anatomical evidence of dyslexia in adults [Klinberg et al. (2000)]. Two groups of children were recruited for the study: a control group consisting of children with normal reading abilities and a case group of children with a previous diagnosis of dyslexia. The subjects were physically and mentally healthy, strongly right-handed, 7–13 years of age, had English as their primary language and intelligence within the average range. There were no significant group differences in age, gender, parental education or socioeconomic status. More details on the study and the image acquisition are given in Deutsch et al. (2005).

The data set consists of 12 diffusion direction maps, of which 6 belong to the control group and 6 to the dyslexic group. Each diffusion direction map is a $95 \times 79 \times 68$ array of voxels representing spatial locations in a rectangular grid with $2 \times 2 \times 3$ mm regular spacings. To every voxel corresponds a unit vector in $\mathbb{R}^3$ that indicates the principal diffusion direction at that voxel. These vectors actually represent axes in the sense that the vectors $\mathbf{x}$ and $-\mathbf{x}$ are equivalent.

For the purposes of statistical analysis, all diffusion direction maps are assumed to be aligned in the same coordinate system so that each voxel corresponds to the same brain structure across all subjects. Since subjects have different head sizes and shapes and may lie slightly differently in the scanner, co-registration of the images was necessary. The common coordinate system used here is a standard called the MNI template. The vertical $z$ axis corresponds to the inferior-superior direction for a subject standing up, looking forward. The origin is located at an anatomical landmark low near the center of the brain called the anterior commissure. In brief, for each subject, both linear and nonlinear transformation parameters were computed from that subject's scalar MRI intensity image by minimizing the square error difference between the transformed image and the template. The DTs were then interpolated entrywise in the transformed grid and principal diffusion directions were recomputed from the DTs at the new locations. The principal diffusion directions were then reoriented by applying the linear portion of the transformation and renormalizing to unit length. For more details on this process, see Schwartzman, Dougherty and Taylor (2005).

Brain investigators often restrict their image analyses to a subset of the brain, called search region or mask, that is relevant to the particular study. The purpose is to increase significance by reducing the data volume and the multiple comparisons problem. A trade-off exists because a search region that is too small will exclude other regions of the brain where interesting differences may be found. Since DTI is particularly good at imaging the white matter of the brain, the search region in this study was defined as voxels that had a high probability of being within the white matter for all subjects [Schwartzman, Dougherty and Taylor (2005)]. In our case, the white matter mask contains $N = 20931$ voxels.

A previous analysis of this dataset [Deutsch et al. (2005)] used scalar FA images instead of the principal diffusion direction maps, and focused on a small white matter region (120 voxels). Our analysis searches for differences in principal diffusion direction over a much larger white matter region (20931 voxels) and reveals differences in gross anatomical structure in other parts of the white matter that are invisible to statistical analyses of FA.

Examples of diffusion direction maps are shown in Figure 1 (these are not maps of individual subjects but rather the average maps for each group, as described in Section 3; the data structure, however, is the same).

## 3. Statistics for diffusion directions.

3.1. *The bipolar Watson distribution.* Given the sign ambiguity property of diffusion directions, it is appropriate to consider probability density functions on the sphere that are antipodally symmetric. If **x** is a random unit vector in $\mathbb{R}^3$, we require that the density $f(\mathbf{x})$ satisfies $f(\mathbf{x}) = f(-\mathbf{x})$. One of the simplest models with this property is the bipolar Watson distribution [Watson (1965)], whose density is given by [Mardia and Jupp (2000), page 181]:

$$(1) \qquad f(\mathbf{x}; \mu, \kappa) = C(\kappa) \exp(\kappa(\mu^T \mathbf{x})^2).$$

The parameter $\mu$ is a unit vector called mean direction and $\kappa$ is a positive constant called concentration parameter. The Watson distribution can be thought of as a symmetrization of the Fisher–Von Mises distribution for unit vectors on the sphere, whose density is $C(\kappa) \exp(\kappa \mu^T \mathbf{x})$. The squared exponent in (1) ensures the required antipodal symmetry. The density has maxima at $\pm\mu$ and becomes more concentrated around $\pm\mu$ as $\kappa$ increases. The density is also rotationally invariant around $\pm\mu$. The normalizing constant $C(\kappa)$ is not needed for the comparison methods used here. It is nevertheless included in the appendix for completeness.

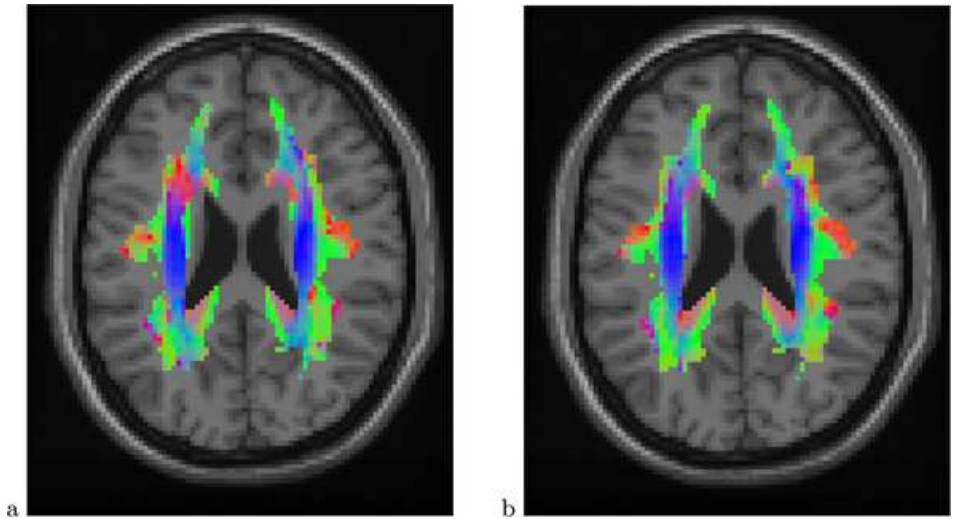

FIG. 1. *Mean diffusion direction map for control group* (a) *and dyslexic group* (b) *at transverse (axial) slice $z = 23$ mm. Colors indicate coordinate directions: superior-inferior (blue), right-left (red) and anterior-posterior (green). The figure was constructed by taking the absolute value of the vector entries of the diffusion direction at each voxel and mapping each (now positive) entry to a scale in the corresponding color. Mixed colors represent directions that are oblique to the coordinate axes. The white matter mask is delineated by the colored area. The gray background is a standard T1-weighted MRI scalar image of the same slice, superimposed for visual reference. Two major brain structures are visible in this picture: the corona radiata (blue vertical stripes on both sides of the brain) contains fibers that run superior-inferior and connect brainstem and cerebellar regions at bottom of the brain with cortical regions at the top of the brain; the corpus callosum (red) connects the left and right brain hemispheres. Notice the difference in direction in the upper left corner of the white matter mask (red in the control group, blue in the dyslexic group).*

More complex models exist for axial data. For instance, the Bingham distribution [Mardia and Jupp (2000), page 181] allows modeling data without assuming rotational invariance about the mean axis. The benefit of flexibility in such models is outweighted by the cost in degrees of freedom for the estimation of the additional parameters. Given the small number of subjects in our data, the simpler Watson model is preferable. More flexible modeling is incorporated instead through the empirical null (Section 4).

3.2. *Mean direction and dispersion.* Let $\mathbf{x}_1, \ldots, \mathbf{x}_n$ be a sample of unsigned random unit vectors in $\mathbb{R}^3$. In our data these would be principal diffusion directions from a single voxel for each of $n$ subjects. Because the data is sign invariant, the direct average is not well defined. Instead, the sample mean direction $\bar{\mathbf{x}}$ is defined as the principal eigenvector (i.e., the eigenvector corresponding to the largest eigenvalue) of the scatter matrix

$$(2) \qquad \mathbf{S} = \frac{1}{n} \sum_{i=1}^{n} \mathbf{x}_i \mathbf{x}_i^T.$$

$S$ may be interpreted as the empirical covariance of the points determined by $\mathbf{x}_1, \ldots, \mathbf{x}_n$ on the sphere. Intuitively, if the points on the sphere have a preferential direction, then, as a group, they are also further apart in space from their antipodes. The principal eigenvector of the scatter matrix points in the direction of maximal variance in space, which is the preferential direction for the points on the sphere. It can be shown (see the Appendix) that $\bar{\mathbf{x}}$ is the maximum likelihood estimator of the location parameter $\mu$ when $\mathbf{x}_1, \ldots, \mathbf{x}_n$ are i.i.d. samples from the Watson model.

The sample dispersion is defined as $s = 1 - \gamma$, where $\gamma$ is the largest eigenvalue of $\mathbf{S}$. Intuitively, when the sample is concentrated around the mean, the antipodes are far apart as a group and so the principal variance $\gamma$ is close to 1, giving a dispersion $s$ that is close to 0. Conversely, when the sample is uniformly scattered on the sphere, the fact that $\text{trace}(\mathbf{S}) = 1$ dictates that the three eigenvalues are equal to $1/3$. The dispersion $s$ in that case takes its maximum value of $2/3$. It can be shown (see the Appendix) that $s$ is the maximum likelihood estimate of $1/\kappa$ in the Watson model, asymptotically when $\kappa \to \infty$. Since $\kappa$ controls concentration, this is consistent with $s$ as a measure of dispersion.

A useful interpretation of $s$ in units of angle is obtained by computing the quantity $\arcsin(\sqrt{s})$, which we call the angle dispersion of the sample. This definition is a direct consequence of the fact (see the Appendix) that $s$ is the average sine-squared of the angles the samples make with the sample mean direction $\bar{\mathbf{x}}$. This definition results in a maximal angle dispersion of $\arcsin(\sqrt{2/3}) = 54.74°$ in the case of uniformity.

An example of mean direction maps is shown in Figure 1. This particular slice was selected via the inference procedure described in Section 4 and it shows large differences in diffusion direction of up to $46.1°$ in the upper left corner of the white matter mask. The statistical test described next formalizes this observation.

3.3. *A two-sample test for directions.* Consider two samples of unsigned unit vectors of sizes $n_1$ and $n_2$ with mean vectors $\mu_1$ and $\mu_2$. We wish to test the null hypothesis $H_0 : |\mu_1 - \mu_2| = 0$ against the alternative $H_A : |\mu_1 - \mu_2| > 0$. The following solution is taken from Mardia and Jupp (2000), page 238,

which assumes equal dispersion in the two groups (akin to a standard $t$-test) and that the samples are highly concentrated around the means. These assumptions are reconsidered in Sections 4.2 and 5.

Under the null, the two samples can be viewed as a single sample of size $n = n_1 + n_2$ and corresponding sample dispersion $s$. Let $s_1$ and $s_2$ denote the sample dispersions of both samples evaluated separately. Similar to an analysis of variance, the total dispersion $ns$ is decomposed as

$$ns = (n_1 s_1 + n_2 s_2) + (ns - n_1 s_1 - n_2 s_2),$$

where the two terms in parenthesis correspond to the intragroup and intergroup dispersion, respectively. The test statistic $T$, which we shall call the Watson statistic, is defined as the ratio of the intergroup to the intragroup dispersion divided by the corresponding number of degrees of freedom, 2 for the intergroup term and $2(n-2)$ for the intragroup term:

$$T = \frac{(ns - n_1 s_1 - n_2 s_2)/2}{(n_1 s_1 + n_2 s_2)/(2(n-2))}. \tag{3}$$

If the underlying concentration parameter $\kappa$ is the same in both samples, then, asymptotically as $\kappa \to \infty$, the Watson statistic (3) has an $F$ distribution with 2 and $2(n-2)$ degrees of freedom. Because of the asymptotic assumptions, this is called a high concentration test rather than a large sample test. This means that the test is valid for small sample sizes as long as the group dispersions are low. The appendix gives a derivation of the null distribution of the Watson statistic in the general case of testing equality of means between a number of samples possibly greater than two.

A map of the Watson statistics is shown in Figure 2a, at the same slice as Figure 1. In our case, $n = 12$ and so the theoretical null distribution is $F(2, 20)$. For reasons that will become clear in Section 4.2, the test statistics have been transformed to a $\chi^2$ scale by a one-to-one quantile transformation from $F(2, 20)$ to $\chi^2(2)$. Notice the local maximum of the test statistic map on the upper left corner of the white matter mask, indicative of the difference in direction alluded in Section 3.2. To assess significance, we incorporate the multiple testing problem, as described next.

## 4. False discovery rate inference.

4.1. *FDR control.* The inference problem of finding significant voxels is a multiple comparisons problem of the type

$$H_0(r) : |\mu_1(r) - \mu_2(r)| = 0 \quad \text{vs.} \quad H_A : |\mu_1(r) - \mu_2(r)| > 0,$$

where the location $r \in \mathbb{R}^3$ ranges over the search region. We overcome the multiple comparisons problem by controlling the false discovery rate (FDR), the expected proportion of false positives among the voxels where the null hypothesis is rejected. As an alternative to the FDR-controlling procedures described in Benjamini and Hochberg (1995) and Genovese, Lazar and Nichols (2002), which are based on ordering of the $p$-values, we use an equivalent interpretation of the procedure taken from Storey, Taylor and Siegmund (2004), as follows.

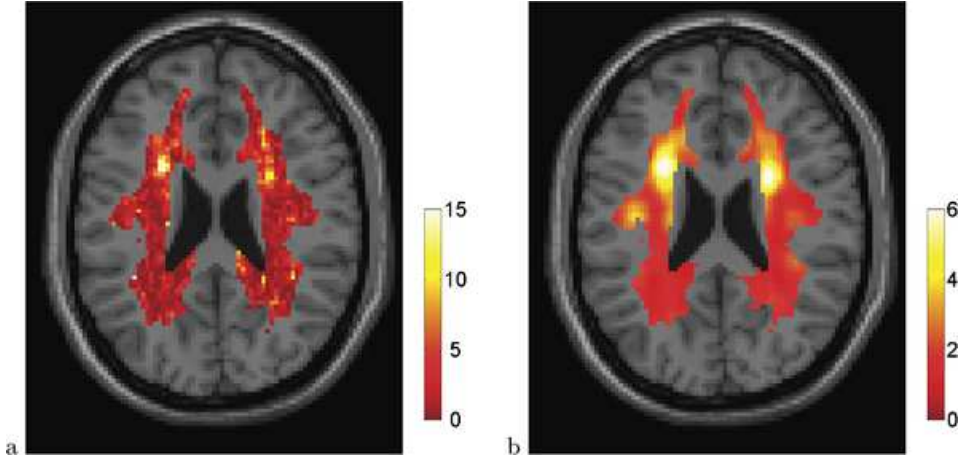

FIG. 2. *Watson statistic map $T(r)$ [after transformation to $\chi^2(2)$]* (a) *and locally averaged map $T_5(r)$ using a $5 \times 5 \times 5$ box smoother* (b), *shown at slice $z = 23$ mm.*

Let $T$ be a test statistic that rejects the null hypothesis at voxel $r$ if its value $T(r)$ is large. In our case, $T$ is the Watson statistic from (3), but the following description applies more generally. Let the search region $M$ contain $N$ voxels, so that $N$ is the number of tests (in our case, $N = 20931$). The null hypothesis is true in an unknown subset $M_0 \subset M$ with $N_0$ voxels, while the alternative is true in the complement $M_A = M \setminus M_0$. The objective is to detect as much as possible of $M_A$ while controlling the FDR. For any fixed threshold $u$, let $R(u)$ and $V(u)$ be respectively the number of rejections and the number of false positives out of $N$. That is,

$$R(u) = \sum_{t \in M} \mathbf{1}(T(r) \geq u), \qquad V(u) = \sum_{t \in M_0} \mathbf{1}(T(r) \geq u).$$

In terms of these empirical processes, the FDR is defined as

$$\mathrm{FDR}(u) = \mathrm{E}\left[\frac{V(u)}{R(u) \vee 1}\right],$$

where the effect of the maximum operator $\vee$ is to set the ratio to 0 when $R(u) = 0$. The natural empirical estimator of $\mathrm{FDR}(u)$ is the ratio

$$(4) \qquad \widehat{\mathrm{FDR}}(u) = \frac{(\hat{N}_0/N) \cdot \mathrm{E}[V(u)/N_0]}{R(u)/N} = \frac{\hat{p}_0 P_{H_0}[T(r) \geq u]}{\hat{P}[T(r) \geq u]},$$

where $\hat{P}$ denotes a probability computed from the empirical distribution of the test statistic $T$ across $M$ and $P_{H_0}$ is a probability computed from the exact distribution of the test statistic according to the null hypothesis $H_0$. The factor $\hat{p}_0$ is an estimate of the true fraction of null voxels $N_0/N$. Assuming that most voxels are null, $\hat{p}_0$ may be taken to be 1, making the estimate $\widehat{\mathrm{FDR}}(u)$ slightly larger and thus conservative. Expression (4) has a nice graphical interpretation as the ratio of the tail areas under the null and empirical densities respectively. Notice that this formula assumes that the null distribution of the test statistic is the same in all voxels.

Voxels in which the alternative hypothesis is true tend to have higher values of the test statistic than expected according to the null hypothesis. As a result, $\widehat{\mathrm{FDR}}(u)$ tends to decrease as $u$ increases. For a given FDR level $\alpha$, the threshold is automatically chosen as the lowest $u$ for which $\widehat{\mathrm{FDR}}(u)$ is smaller or equal to $\alpha$:

$$u_\alpha = \inf\{u : \widehat{\mathrm{FDR}}(u) \leq \alpha\}.$$

It is shown by Storey, Taylor and Siegmund (2004) that when the truly null $N_0$ test statistics are independent and identically distributed, this procedure (with $\hat{p}_0 = 1$) is equivalent to the Benjamini and Hochberg (1995) procedure, and therefore provides strong control of the FDR. Moreover, it is shown in Storey, Taylor and Siegmund (2004) that the strong control also holds asymptotically for large $N$ under weak dependence of the test statistics, such as dependence in finite blocks. Weak dependence may be assumed in brain imaging data because the number of voxels is large and dependence is usually local with an effective range that is small compared to the size of the brain.

4.2. *Empirical null.* A histogram of the Watson statistics for all $N = 20931$ voxels in the white matter mask is shown in Figure 3a, except that the test statistics have been transformed to a $\chi^2$ scale by a one-to-one quantile transformation from $F(2, 20)$ to $\chi^2(2)$. The theoretical null $\chi^2(2)$ (dashed curve) gives a reasonable description of the distribution of the test statistics. The empirical null density (solid curve), however, provides a much better fit to the data. The empirical null takes advantage of the large number of voxels to globally correct for the lack of flexibility and possibly short-of-asymptotic behavior of the distribution prescribed by the theoretical model.

The empirical null concept was originally proposed for $z$-scores [Efron (2004)], whose theoretical null is $N(0, 1)$. There, $t$-statistics were handled by transforming them to $z$-scores via a one-to-one quantile transformation from the appropriate $t$ distribution to a $N(0, 1)$. The effect of this transformation is to eliminate the dependence on the number of subjects, which affects the estimation of the variance in the denominator of the $t$-statistic.

In our case, the transformation from $F(2,20)$ to $\chi^2(2)$ has a similar effect. Keeping the numerator degrees of freedom intact preserves interpretation of the dimensionality of the problem. The empirical null for $\chi^2$ statistics is computed as follows [Schwartzman (2006)].

Let $f(t)$ denote the marginal density of the test statistic $T$ over all voxels. From the setup of Section 4.1, we write it as the mixture

$$f(t) = p_0 f_0(t) + (1-p_0) f_A(t),$$

where the fraction of voxels $p_0 = N_0/N$ behave according to the null density $f_0(t)$ and the remainder $1-p_0$ behave according to an alternative density $f_A(t)$. To make the problem identifiable, it is assumed that $p_0$ is close to 1 (say, larger than 0.9), so that the bulk of the histogram $N\Delta \hat{f}(t)$ (left portion of Figure 3a, where $\Delta = 0.2$ is the bin width) is mostly composed of null voxels. The density $f_A(t)$ may itself be a mixture of other components but its form is irrelevant as long as it has most of its mass away from zero, or equivalently, if its contribution to the mixture $(1 - p_0)f_A(t)$ is small for values of $t$ close to zero.

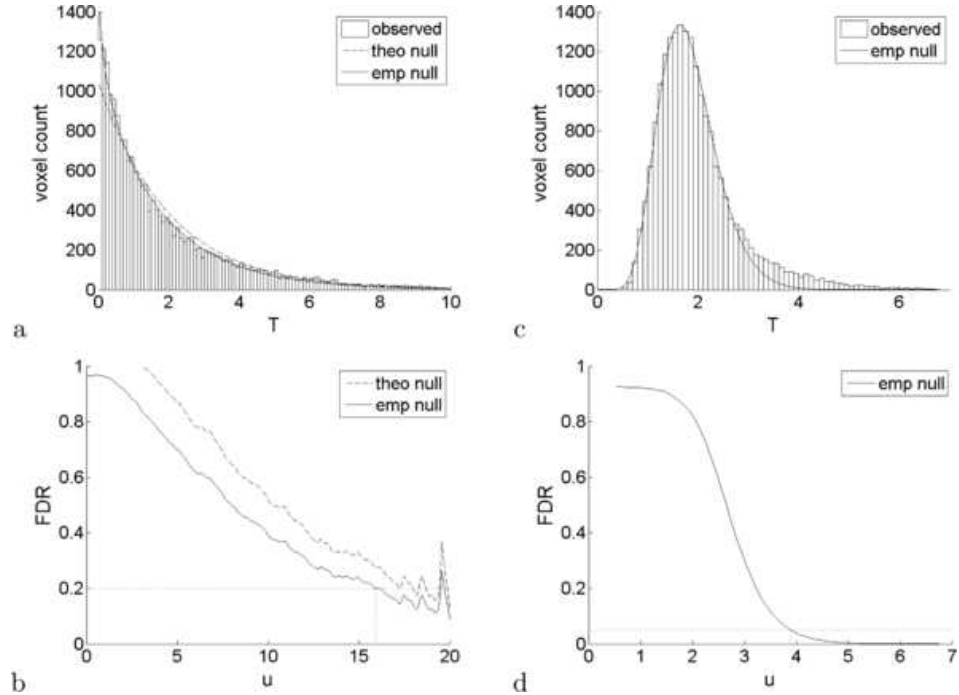

FIG. 3. *FDR analysis with no smoothing (left column) and smoothing using $b=5$ (right column). Top row: observed histogram of the $T$ statistic [after transformation to $\chi^2(2)$] compared to the theoretical null density (dashed curve) and empirical null density (solid curve). Bottom row: Estimates of FDR as a function of the threshold $u$ according to the theoretical null (dashed curve) and empirical null (solid curve). No theoretical null is available for the smoothed test statistic (right column).*

As an adjustment to the theoretical null $N(0, 1)$, Efron (2004) proposed an empirical null of the form $N(\mu, \sigma^2)$. Similarly, as an adjustment to a theoretical null $\chi^2(\nu_0)$ with $\nu_0$ degrees of freedom, we propose an empirical null of the form $a\chi^2(\nu)$ with $\nu$ degrees of freedom (possibly different from $\nu_0$) and scaling factor $a$ (possibly different from 1), that is,

$$(5) \qquad f_0(t) = \frac{1}{(2a)^{\nu/2}\Gamma(\nu/2)} e^{-t/(2a)} t^{\nu/2 - 1}.$$

This is essentially a gamma distribution, but the scaled $\chi^2$ notation makes interpretation of the results easier. Under the above assumptions, the portion of the histogram $N\Delta \hat{f}(t)$ close to $t = 0$ should resemble the scaled null $p_0 f_0(t)$. Proceeding as in Efron (2004), we fit model (5) to the histogram $N\Delta \hat{f}(t)$ via Poisson regression using the link

$$(6) \qquad \log(p_0 f_0(t)) = -\frac{t}{2a} + \left(\frac{\nu}{2} - 1\right) \log t + \text{constant}.$$

This is a linear model with predictors $t$ and $\log t$ and observations given by the histogram counts. The estimated parameters $\hat{a}$ and $\hat{\nu}$ are solved from the estimated coefficients of $t$ and $\log t$ in the Poisson regression. An estimate of $p_0$ is also obtained by solving the expression for the constant in the regression.

As in Efron (2004), the fitting interval is arbitrary. In Figure 3a we used an interval from 0 up to the 90th percentile of the histogram. The fitted parameters were $\hat{a} = 1.000$ and $\hat{\nu} = 1.78$. Although the scaling is unaffected, the reduced number of degrees of freedom may be capturing some additional structure not accounted for by the Watson model, such as correlation or spherical asymmetry.

With the empirical null, the FDR estimate (4) is now replaced by

$$(7) \qquad \widehat{\text{FDR}}^+(u) = \frac{\hat{p}_0 \hat{P}_{H_0}[T(r) \geq u]}{\hat{P}[T(r) \geq u]}.$$

Notice the extra "hat" in the numerator, indicating that the empirical null is being used instead of the theoretical null.

The FDR analysis is summarized in Figure 3b. The FDR curve corresponding to the theoretical null (dashed) was computed using (4) with $\hat{p}_0 = 1$. The FDR curve corresponding to the empirical null curve (solid) was computed using (7) with the fitted parameters described above. The value of the curve for $u = 0$ is our empirical null estimate of $p_0$, equal to 0.974. Notice that the FDR curves have a general tendency to decrease as the threshold increases. The empirical null gives better FDR values than the theoretical null, but that is not necessarily true in general [Efron (2004) provides a counterexample].

The FDR level $\alpha = 0.2$ intersects the empirical null FDR curve at a threshold of 15.92, marked in the figure as a vertical dashed segment. As a reference, this threshold corresponds to an uncorrected $p$-value of $3.5 \times 10^{-4}$. The 15.92 threshold resulted in 23 interesting voxels. Although these selected voxels are located in several areas of the white matter, it is in slices $z = 23$ mm to $z = 25$ mm that they are closer together and have the highest values of the Watson statistic. These three slices are shown in the top row of Figure 4, with the corresponding subset of 8 voxels marked in white. The group difference in this region can also be seen as a local maximum in the test statistic map of Figure 2a. A hierarchical clustering analysis of the selected voxels showed that the 23 voxels can be grouped into 14 clusters, the largest of which has size 3. These results are also indicated in Table 1.

4.3. *Improved power by local averaging.* So far, the analysis has been based on marginal densities and has not taken into account the information available in the location index $r$. Neighboring voxels tend to be similar because the anatomical structures visible in DTI are typically larger than the voxel size. The logical spatial units are the various brain structures, not the arbitrary sampling grid of voxels, and therefore, it is desirable to select clusters rather than individual voxels. Spatial smoothing may reduce noise and may better detect clusters that correspond to actual anatomical structures.

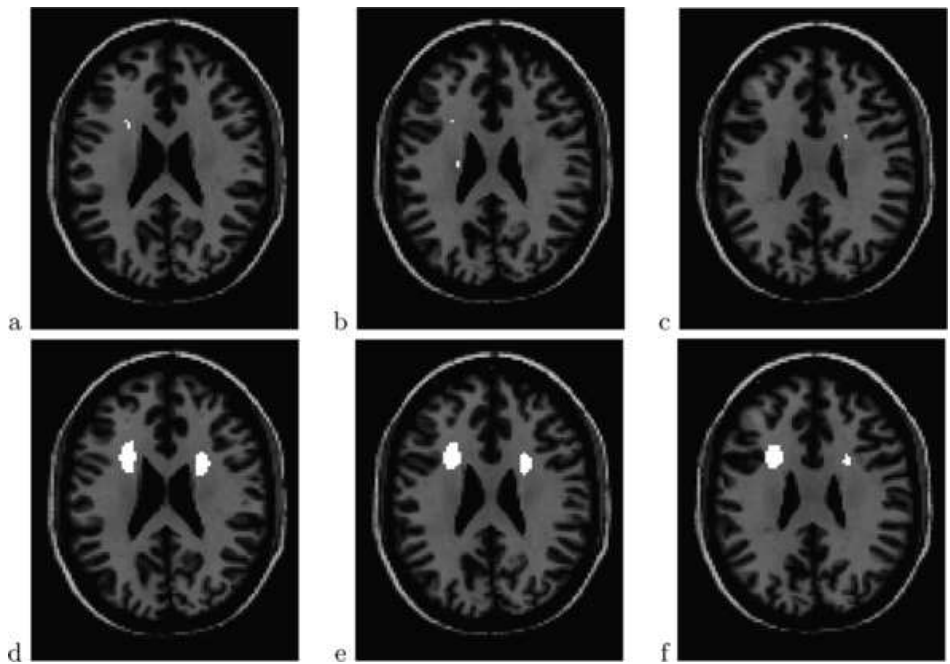

FIG. 4. *Interesting voxels (white) thresholded from unsmoothed test statistics at FDR level 0.2 (top row) and using kernel size $b = 5$ at FDR level 0.05 (bottom row). Both analyses are based on the empirical null. Shown slices are $z = 23$ mm (left column), 25 mm (middle column) and 27 mm (right column).*

TABLE 1
*Interesting voxels selected at various smoothing sizes b and FDR levels $\alpha$. Listed are the mask size $N$, the estimated empirical null parameters, fitting limit $T_{90}$, FDR level $\alpha$, threshold $u_\alpha$, number of selected voxels $R(u_\alpha)$, number of clusters and size of the largest clusters. The $\leq$ sign implies there are other smaller clusters than listed*

| $b$ | $N$ | $\hat{p}_0$ | $\hat{a}$ | $\hat{\nu}$ | $T_{90}$ | $\alpha$ | $u_\alpha$ | $R(u_\alpha)$ | # clust | clust sz |
|---|---|---|---|---|---|---|---|---|---|---|
| 1 | 20931 | 0.974 | 1.000 | 1.78 | 4.84 | 0.2 | 15.92 | 23 | 14 | $\leq$1,2,3 |
| 3 | 20613 | 0.938 | 0.203 | 8.66 | 3.51 | 0.2 | 4.14 | 1273 | 53 | $\leq$59,282,478 |
|   |   |   |   |   |   | 0.05 | 5.46 | 452 | 25 | $\leq$34,89,192 |
|   |   |   |   |   |   | 0.01 | 6.80 | 164 | 13 | $\leq$21,36,57 |
| 5 | 19856 | 0.928 | 0.091 | 20.01 | 3.03 | 0.2 | 3.23 | 1609 | 25 | $\leq$90,427,711 |
|   |   |   |   |   |   | 0.05 | 3.90 | 790 | 21 | $\leq$42,216,431 |
|   |   |   |   |   |   | 0.01 | 4.60 | 345 | 10 | $\leq$33,86,167 |
| 7 | 18720 | 0.926 | 0.052 | 36.10 | 2.80 | 0.2 | 2.90 | 1606 | 18 | $\leq$83,509,889 |
|   |   |   |   |   |   | 0.05 | 3.35 | 829 | 12 | $\leq$17,256,517 |
|   |   |   |   |   |   | 0.01 | 3.79 | 442 | 8 | $\leq$13,108,316 |
| 9 | 17050 | 0.930 | 0.035 | 54.31 | 2.66 | 0.2 | 2.76 | 1410 | 12 | $\leq$11,519,852 |
|   |   |   |   |   |   | 0.05 | 3.13 | 691 | 6 | $\leq$3,187,494 |
|   |   |   |   |   |   | 0.01 | 3.52 | 227 | 6 | $\leq$4,216 |

Consider a simple box smoother $h_b(r) = \mathbf{1}(r \in B_b)/|B_b|$, where the box $B_b$ is a cube of side $b$ voxels and volume $|B_b| = b^3$. Convolution of the test statistic map $T(r)$ with the box smoother results in the locally averaged test statistic map

$$(8) \qquad T_b(r) = T(r) * h_b(r) = \frac{1}{|B_b|} \sum_{v \in B_b} T(r-v).$$

In the null regions, the smoothed test statistic $T_b(r)$ at every voxel is the average of $b^3$ $\chi^2(2)$-variables. If the test statistics were independent, $T_b(r)$ would be exactly $\chi^2(2b^3)/b^3$. It is known that the sum of identically distributed exponentially correlated gamma-variables can be well approximated by another gamma-variable [Kotz and Adams (1964)]. Instead of theoretically deriving the parameters of the gamma distribution and estimating the correlation from the data, an easier solution is given by the empirical null. Exponential correlation being a reasonable model for spatial data, we take the empirical null to be a scaled $\chi^2$ (gamma). We then estimate the null parameters $a$ and $\nu$ directly from the histogram of $T_b(r)$ following the same recipe as in Section 4.2.

A slight change from the previous analysis of Section 4.2 is in the size of the mask. In order to minimize edge effects, the local averaging (8) was applied to the unmasked images and the mask reapplied for the purposes of statistical analysis. Edge effects due to some external anatomical features close to our search region (such as cerebro-spinal fluid) resulted in the exclusion of some voxels, causing a slight reduction in the size of the mask from the original $N = 20931$ at $b = 1$ to $N = 17050$ at $b = 9$ (Table 1).

The locally averaged test statistic map for $b = 5$ is shown in Figure 2b. The corresponding histogram is shown in Figure 3c. Notice the narrowing of the histogram around the global mean value 2 as a result of the averaging. Figure 3d shows the corresponding FDR curve. The smoothing has greatly helped differentiate the two major components of the mixture. This time we can afford to reduce the inference level substantially with respect to the previous analysis. Setting $\alpha = 0.05$ results in 1609 interesting voxels out of 19856. Figure 4, bottom row, shows the largest two out of 21 discovered clusters. These clusters actually extend vertically in the brain all the way from $z = 5$ mm to $z = 27$ mm.

Repeated analyses for $b = 3, 5, 7, 9$ result in the selection of two large clusters in about the same location. The effect of $b$ on the number of selected voxels and cluster size is summarized in Figure 5. The total fraction of selected voxels out of $N$ increases dramatically even with the least amount of smoothing, but it reaches a maximum at $b = 7$. A similar plateau effect is also observed in the size of the two largest clusters, especially at FDR levels 0.05 and 0.01. Notice that the second largest cluster disappears at $b = 9$ and FDR level 0.01. Increasing $b$ beyond 9 is impractical due to the limited size of the white matter mask.

**5. Discussion.** We have compared two groups of diffusion direction maps using a Watson model for directional data. The inference procedure was built upon voxelwise test statistics and depended only on their marginal distribution. Taking advantage of the large number of voxels and the spatial structure of the data, we were able to improve the model fit to the data using global parameters and improve the statistical power using local averaging.

The choice of the null distribution is crucial for the inference process. Why is the theoretical null not enough? The $F(2, 20)$ theoretical null (Section 3.3) is a high concentration asymptotic based on a normal approximation to the Watson density (see the Appendix). The asymptotic density is actually approached quickly as $\kappa$ increases (Figure 6). For example, the 0.001-quantile is 8.5 for $\kappa = 5$ and 9.4 for $\kappa = 10$, compared to 9.9 for the $F(2, 20)$ distribution. Since the 25th and 50th percentiles of the distribution of the estimated $\kappa$ among all 9203 voxels are 5.0 and 9.8, we may say the high concentration assumption is reasonable for many voxels, yet the minority that is not highly concentrated may have an effect on the overall mixture.

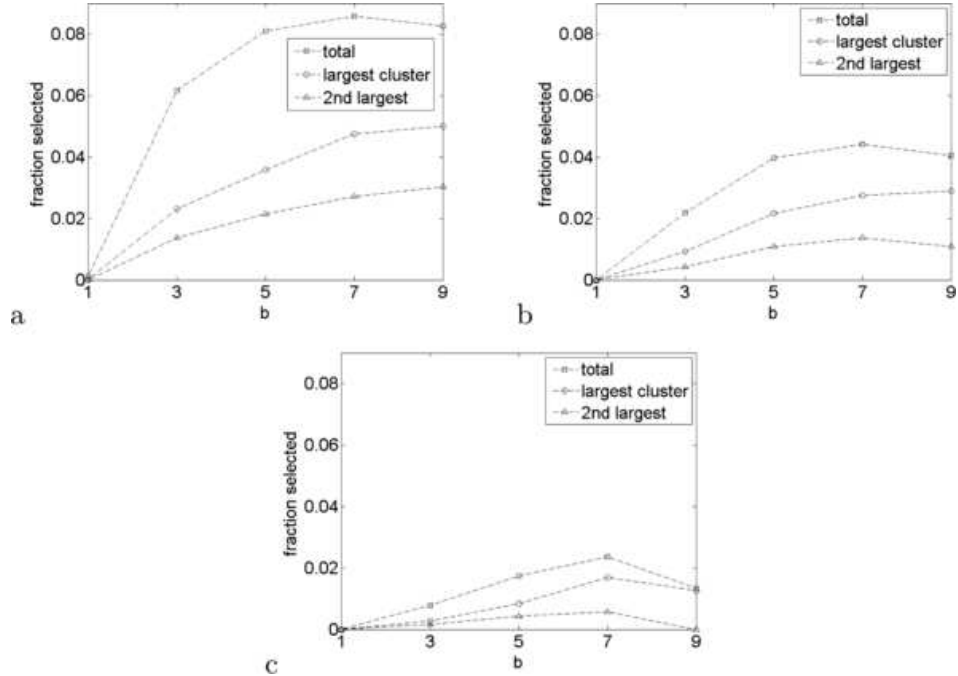

FIG. 5. *Selected voxels as a fraction of total mask size N for FDR levels 0.2* (a), *0.05* (b) *and 0.01* (c). *Indicated are the total set of selected voxels and the largest two clusters.*

Although not obvious from Figure 6, the $F(2,20)$ density is heavier tailed than for finite $\kappa$. Notice in the above calculation that the 0.001-quantile 9.9 from the $F(2,20)$ density is higher than it would be if the true con-centration were used instead. Since the Watson density has a finite domain on the sphere, it is necessarily lighter tailed than the normal density that approximates it when $\kappa$ is large. This effect is stronger in the numerator of the Watson statistic (3), thus making the $F$ distribution heavier tailed than necessary for the data. The empirical null provides the lighter tail for finite $\kappa$, as needed.

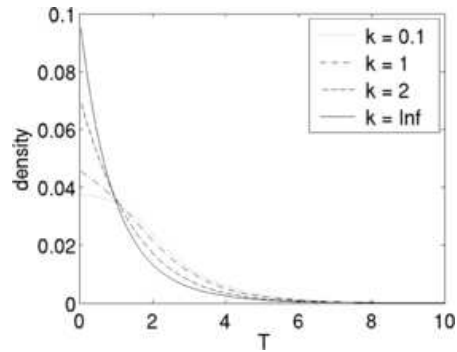

FIG. 6. *Simulated null densities for various values of the concentration parameter $\kappa$. The high-concentration asymptotic $F(2,20)$ is labeled as $\kappa = \infty$.*

The discrepancy between the theoretical and empirical nulls for low values of $T$ may be explained by the distribution on the sphere not being spherically symmetric. The numerator degrees of freedom in $F(2, 20)$ corresponds to the dimensionality of the normal approximation to the Watson density on the tangent plane when $\kappa$ is large. The number of degrees of freedom 1.78 in the empirical null, somewhat smaller than 2, suggests that the proper approximation may not be bivariate normal with circular contours but rather with elliptical contours. The change in number of degrees of freedom may also be a consequence of unequal dispersions between the two groups, akin to the scalar case [Scheffé (1970)]. Again, instead of paying extra parameters at each voxel, this is captured globally by the empirical null. The empirical null is also effective because it provides a model for a mixture of distributions from a large number of voxels, adjusting for unknown heteroscedasticity and correlation between individual voxels.

As additional null validation, an analysis was performed in which half the subjects were swapped between the two groups in order to remove the group effect. The histogram and fitted empirical null were very similar to the ones obtained for the original groups. Surprisingly, however, some significant voxels were found at the same FDR levels as the original group comparison, although substantially less in number. This may be an indication that the empirical null, while it helps supplement the deficiencies of the theoretical model, may still not be enough to explain all the null variation in the data.

It should be noted that, in general, there is not necessarily a direct increase in power associated with the empirical null [Efron (2004) provides counterexamples]. The empirical null only answers a question of model validity. An alternative option to the empirical null would be to do a permutation test with the same Watson statistic. In our case the permutation test has little power because the lowest $p$-value attainable with two groups of 6 subjects is 0.001, which cannot survive the multiple comparisons problem with 20931 voxels.

Local averaging has a tremendous impact on statistical power because the power at every single voxel is indeed low. Consider the power at a single voxel with the observed peak separation of $46.1°$ as the effect size. Simulation of the Watson statistic under this alternative hypothesis reveals that the power of a single test of the $F(2, 20)$ null at level $\alpha = 0.001$ is 0.180 for $\kappa = 5$ and 0.804 for $\kappa = 10$. A very high concentration is required in order to have sufficient power. Under the assumption that the signal changes slowly over space, local averaging has the effect of reducing variance, effectively increasing the concentration associated with the smoothed test statistic and thus increasing power.

The reduction in variance provided by local averaging increases with the size of the smoothing kernel. Too much smoothing, however, results in a reduction of the signal. This effect is seen in Figure 5, where the detection rate goes down if the kernel size $b$ is too large. The shape of the graphs in Figure 5 suggests there may exist an optimal kernel that maximizes power, although the exact choice of $b$ might not be critical as long as it is within a certain neighborhood of the optimum. The choice of kernel size is an interesting question for further research and prompts questions about the proper definition of power for FDR inference of spatial signals.

Another interesting effect is seen in the behavior of the estimated parameters of the empirical null as a function of the kernel size $b$ (Figure 7). While $\hat{p}_0$ resembles Figure 5, close inspection reveals that $\hat{a}$ and $\hat{\nu}$ are very close to functional forms of $b$, respectively $1/\sqrt{b^3}$ and $2\sqrt{b^3}$. These rates are slower than the rates $1/b^3$ and $b^3$ we would expect if neighboring voxels were independent. The actual rates might shed light into the correlation structure of the data.

Despite the apparent success of smoothing, there are some caveats. As $b$ increases, so does the dependence between the test statistics, shaking the ground on which the strong control of FDR relies (Section 4.1). Furthermore, there is a problem of interpretation of the results. The inference after smoothing is no longer about the original set of hypotheses but about a smoothed set of hypotheses. We might gain significance, but loose spatial localization. Also anatomically, smoothing with a kernel larger than the structures of interest will prevent interpretation of the results as differences in brain structure.

The choice of smoothing the test statistic map, as opposed to the original data, was a practical one. To smooth the subjects' direction maps was problematic because, even if the Watson model were correct, the mean direction of a Watson sample is no longer Watson. Smoothing of the test statistic map is a more general platform that can be studied independently of how the test statistics were generated.

While significant differences were found between the two groups of principal diffusion direction images, the results should be taken with some caution, as in any other observational study. At the core of the voxelwise comparison paradigm used in this study is the difficulty to tell how much of the effect is anatomical and how much is due to the image alignment process. Not enough is known yet about the anatomical basis for dyslexia in order to interpret the results. To gain insights about dyslexia would require tracking the neural connections between the deep white matter regions discovered in this study and the peripheral gray matter regions involved in reading. This is a challenge beyond the scope of this paper.

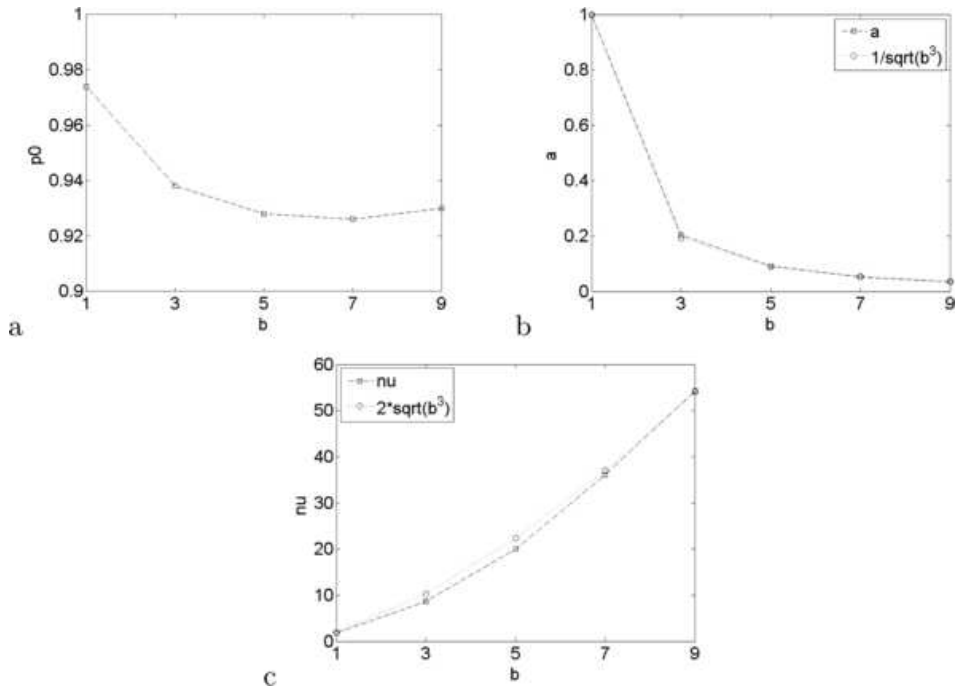

Fig. 7. *Empirical null parameters as a function of kernel size $b$: $\hat{p}_0$ (a), $\hat{a}$ (b) and $\hat{\nu}$ (c). Both $\hat{a}$ and $\hat{\nu}$ resemble explicit functional forms of $b$.*

In summary, we have developed a methodology for comparing two groups of diffusion direction maps and finding interesting regions of difference between the two groups. The Watson model was necessary because of the directional nature of the data. The inference procedure, on the other hand, was built upon voxelwise test statistics. The key elements in the inference procedure were the empirical null and smoothing of the test statistic map. The procedure can be applied more generally to other large scale simultaneous hypothesis testing problems with a continuous underlying spatial structure. The requirements are an approximate theoretical null distribution of the test statistics, upon which an empirical null distribution can be computed, and a spatial structure where spatial smoothing of the test statistics is well defined.

## APPENDIX: COMPUTATIONS FOR THE BIPOLAR WATSON DISTRIBUTION

The following summary is a reinterpretation of material from Mardia and Jupp (2000), pages 181, 202, 236–240, Watson (1965) and Best and Fisher (1986).

It includes a new asymptotic approximation for the integration constant and a new interpretative quantity called angle dispersion.

**A.1. Integration constant.** Define a spherical coordinate system on the unit sphere so that the $z$-axis coincides with the mean vector $\mu$. For a unit vector $\mathbf{x}$, let $\theta$ be the co-latitude angle between $\mathbf{x}$ and the $z$-axis. Denote the longitude angle by $\phi$. The Watson density in this coordinate system is given by

$$f(\theta, \phi) = C(\kappa) e^{\kappa \cos^2 \theta} \sin \theta \, d\theta \, d\phi, \qquad 0 \leq \theta < \pi/2, 0 \leq \phi < 2\pi.$$

The restriction of the density to half the sphere accounts for the antipodal symmetry. This formulation is slightly different from the one in Best and Fisher (1986), which defines the density on the entire sphere. An expression for $C(\kappa)$ is obtained integrating the density with the change of variable $u = \cos \theta$, yielding

$$C(\kappa) = \left[ 2\pi \int_0^1 e^{\kappa u^2} \, du \right]^{-1}.$$

The definite integral in the above expression is a special case of the Dawson integral [Abramowitz and Stegun (1972), page 298].

An explicit asymptotic expression can be found in the large concentration case. When $\kappa$ is large, most of the probability density is concentrated around $\mu$ and $\mathbf{x}$ is close to $\mu$ with high probability. Intuitively, the region of the sphere close to $\mu$ looks locally like a two-dimensional plane. A scaled projection of the density onto this plane is obtained with the change of variable $r = \sqrt{2\kappa} \sin \theta$, giving

$$(9) \quad f(r, \phi) = \frac{2\pi C(\kappa) e^\kappa}{2\kappa} \cdot \frac{e^{-r^2/2} r \, dr \, d\phi}{2\pi \sqrt{1 - r^2/2\kappa}}, \qquad 0 \leq r < \sqrt{2\kappa}, 0 \leq \phi < 2\pi.$$

For large $\kappa$ the second factor in the density looks like a bivariate Gaussian density and its integral should converge to 1. Indeed, another change of variable $u = r^2/2\kappa$ and integrating by parts, the second factor in (9) yields

$$\int_0^{2\pi} \frac{d\phi}{2\pi} \int_0^1 \frac{e^{-\kappa u} \kappa \, du}{\sqrt{1 - u}} = 2\kappa \left( 1 - \kappa \int_0^1 e^{-\kappa u} \, du \sqrt{1 - u} \right).$$

The bounds

$$1 - \frac{u}{2} - \frac{u^2}{2} \leq \sqrt{1 - u} \leq 1 - \frac{u}{2}, \qquad 0 \leq u \leq 1,$$

then lead to

$$1 + (\kappa - 1)e^{-\kappa} \leq \int_0^{\sqrt{2\kappa}} \frac{e^{-r^2/2} r \, dr}{\sqrt{1 - r^2/2\kappa}} \leq 1 + \frac{2}{\kappa} - \left( 3 + \frac{2}{\kappa} \right) e^{-\kappa}.$$

Replacing in the integral of (9), we obtain

$$\text{(10)} \qquad \frac{\pi C(\kappa) e^\kappa}{\kappa} \sim 1 \;\Rightarrow\; C(\kappa) \sim \frac{\kappa}{\pi e^\kappa}, \qquad \kappa \to \infty.$$

**A.2. Maximum likelihood estimates.** Let $\mathbf{x}_1, \ldots, \mathbf{x}_N$ be a random sample from the Watson distribution. The log-likelihood is

$$\text{(11)} \qquad \kappa \sum_{i=1}^N (\mathbf{x}_i^T \mu)^2 + N \log C(\kappa) = N\{\kappa \mu^T \mathbf{S} \mu + \log C(\kappa)\},$$

where $\mathbf{S}$ is the scatter matrix 2. For $\kappa > 0$, the MLE $\hat{\mu}$ is the maximizer of $\mu^T \mathbf{S} \mu$ constrained to $\mu^T \mu = 1$ and is given by the eigenvector of $\mathbf{S}$ that corresponds to the largest eigenvalue $\gamma$. At the maximum,

$$\text{(12)} \qquad \hat{\mu}^T \mathbf{S} \hat{\mu} = \hat{\mu}^T \gamma \hat{\mu} = \gamma.$$

Differentiation of (11) with respect to $\kappa$ gives $\hat{\mu}^T \mathbf{S} \hat{\mu} = A(\hat{\kappa})$, where

$$\text{(13)} \qquad A(\kappa) = -\frac{C'(\kappa)}{C(\kappa)} = \frac{\int_0^1 t^2 e^{\kappa t^2}\, dt}{\int_0^1 e^{\kappa t^2}\, dt}.$$

Using (12), $\hat{\kappa}$ is thus found by solving

$$\text{(14)} \qquad A(\hat{\kappa}) = \gamma.$$

The function $A(\kappa)$ is monotonically increasing in the range $[1/3, 1)$ as $\kappa$ increases from 0 to $\infty$. Replacing the asymptotic (10) in (13), we obtain the large concentration approximation

$$A(\kappa) \sim 1 - \frac{1}{\kappa}, \qquad \kappa \to \infty.$$

Setting the dispersion $s = 1 - \gamma$ in (14) and using the previous approximation for $A(\kappa)$, we get that at the point of maximum likelihood $s \sim 1/\hat{\kappa}$, which justifies the interpretation of $s$ as a measure of dispersion.

We now obtain an interpretation of $s$ in terms of angle units. Replacing (2) in (12), we obtain

$$\gamma = \hat{\mu}^T \mathbf{S} \hat{\mu} = \frac{1}{N} \sum_{i=1}^N (\hat{\mu}^T \mathbf{x}_i)(\hat{\mu}^T \mathbf{x}_i)^T = \frac{1}{N} \sum_{i=1}^N \cos^2 \hat{\theta}_i,$$

and so

$$\text{(15)} \qquad s = 1 - \gamma = \frac{1}{N} \sum_{i=1}^N \sin^2 \hat{\theta}_i.$$

In other words, $s$ is the average sine-squared of the angles that the samples make with the mean direction. An interpretation of $s$ in units of angle is obtained thus by computing the quantity $\arcsin(\sqrt{s})$, which we call angle dispersion.

**A.3. A multi-sample large concentration test.** Given $q$ samples of sizes $N_1, \ldots, N_q$, we wish to test $H_0 : \mu_1 = \cdots = \mu_q$ against the alternative that at least one of the means is different. For simplicity, we assume that all samples have the same unknown large concentration $\kappa$.

Consider first the entire sample of size $N = N_1 + \cdots + N_q$ with common mean $\mu$ and pooled dispersion $s$. Using (15), we write the total dispersion as

$$2\kappa N s = \sum_{i=1}^{N} 2\kappa \sin^2 \hat{\theta}_i = \sum_{i=1}^{N} \hat{r}_i^2,$$

where $\hat{r}_i = \sqrt{2\kappa} \sin \hat{\theta}_i$. When $\mu$ is known, the density (9) indicates that each $r_i$ (without the "hat") is approximately standard bivariate Gaussian when $\kappa$ is large. $2\kappa N s$ is thus the sum of $N$ independent approximately $\chi_2^2$ random variables. The estimation of $\mu$ reduces two degrees of freedom so

$$(16) \qquad 2\kappa N s \underset{H_0}{\overset{\cdot}{\sim}} \chi_{2(N-1)}^2.$$

For $q$ independent samples of sizes $N_1, \ldots, N_q$ and dispersions $s_1, \ldots, s_q$, $2q$ parameters are fitted and we have the intragroup sum of squares

$$(17) \qquad 2\kappa \sum_{j=0}^{q} N_j s_j \underset{H_0}{\overset{\cdot}{\sim}} \chi_{2(N-q)}^2.$$

In the "analysis of variance" decomposition

$$2\kappa N s = 2\kappa \sum_{j=0}^{q} N_j s_j + 2\kappa \left[ N s - \sum_{j=0}^{q} N_j s_j \right],$$

the asymptotics (16) and (17) imply that the second term on the RHS is approximately $\chi^2$ with $2(N-1) - 2(N-q) = 2(q-1)$ degrees of freedom and approximately independent of the first term. The second term represents the intergroup dispersion. Proceeding as in the analysis of variance for normal variables, we construct the Watson test statistic as the ratio between the intergroup and the intragroup terms divided by the appropriate number of degrees of freedom. Correspondingly, the Watson statistic is asymptotically $F$-distributed as

$$T = \frac{[Ns - \sum_{j=0}^{q} N_j s_j]/2(q-1)}{[\sum_{j=0}^{q} N_j s_j]/2(N-q)} \underset{H_0}{\overset{\cdot}{\sim}} F_{2(q-1), 2(N-q)}.$$

Notice that the actual value of $\kappa$, although assumed large, is not needed in the final formula of the statistic.

A. Schwartzman
Department of Biostatistics
Dana-Farber Cancer Institute
Mailstop LW225
44 Binney Street
Boston, Massachusetts 02115
USA
E-mail: armins@hsph.harvard.edu

R. F. Dougherty
Department of Psychology—Jordan Hall
450 Serra Mall
Stanford, California 94305-2130
USA
E-mail: bobd@stanford.edu

J. E. Taylor
Department of Statistics—Sequoia Hall
390 Serra Mall
Stanford, California 94305-4065
USA
E-mail: jonathan.taylor@stanford.edu